\documentclass[apj]{emulateapj}
\usepackage{graphics}
\usepackage{psfig}
\usepackage{epsfig}
\usepackage{alltt}

\slugcomment{\it Submitted for Publication in ApJ}
\shortauthors{Comerford, Meneghetti, Bartelmann, \& Schirmer}
\shorttitle{Mass Distributions of Clusters from Arcs}

\begin{document}

\title{Mass Distributions of HST Galaxy Clusters from Gravitational Arcs}

\author{Julia M. Comerford\altaffilmark{1}, Massimo
Meneghetti\altaffilmark{2}, Matthias Bartelmann\altaffilmark{2}, and
Mischa Schirmer\altaffilmark{3}}

\affil{$^1$Astronomy Department, 601 Campbell Hall, University of
California, Berkeley, CA 94720}
\affil{$^2$Zentrum f\"ur Astronomie, Institut f\"ur Theoretische
  Astrophysik, Universit\"{a}t Heidelberg, Albert-\"Uberle-Str. 2,
  69120 Heidelberg, Germany}
\affil{$^3$Isaac Newton Group of Telescopes, Calle Alvarez Abreu 68,
38700 Santa Cruz de La Palma, Tenerife, Spain}

\begin{abstract}
Although $N$-body simulations of cosmic structure formation suggest
that dark matter halos have density profiles shallower than isothermal
at small radii and steeper at large radii, whether observed galaxy
clusters follow this profile is still ambiguous. We use one such
density profile, the asymmetric NFW profile, to model the mass
distributions of 11 galaxy clusters with gravitational arcs observed
by HST. We characterize the galaxy lenses in each cluster as NFW
ellipsoids, each defined by an unknown scale convergence, scale
radius, ellipticity, and position angle. For a given set of values of
these parameters, we compute the arcs that would be produced by such a
lens system. To define the goodness of fit to the observed arc system,
we define a $\chi^2$ function encompassing the overlap between the
observed and reproduced arcs as well as the agreement between the
predicted arc sources and the observational constraints on the source
system. We minimize this $\chi^2$ to find the values of the lens
parameters that best reproduce the observed arc system in a given
cluster. Here we report our best-fit lens parameters and corresponding
mass estimates for each of the 11 lensing clusters. We find that
cluster mass models based on lensing galaxies defined as NFW
ellipsoids can accurately reproduce the observed arcs, and that the
best-fit parameters to such a model fall within the reasonable ranges
defined by simulations. These results assert NFW profiles as an
effective model for the mass distributions of observed clusters.
\end{abstract}


\section{Introduction}
\label{intro}

While numerical simulations of dark-matter halos in the CDM model of
cosmic structure formation invariably predict density profiles which
are steeper than isothermal outside and flatter inside a scale radius
which is of order 20\% the virial radius for cluster-sized halos
(e.g.\ \citealt{NA97.1,MO98.1,PO03.1,NA04.1}), it is yet unclear
whether real galaxy clusters have such density profiles. Galaxy
rotation curves (see \citealt{SO01.1} for a review) and strong-lensing
constraints (e.g.\ \citealt{RU01.1,RU03.1,TR04.1,KE01.1}) have shown
that galaxies need to have at least approximately isothermal density
profiles which are however the result of baryonic physics such as gas
cooling and star formation. In galaxy clusters, baryonic effects
should be substantially weaker, and thus their density profiles
outside the innermost cores should still reflect the typical CDM
density profile found in numerical simulations.

Gravitational lensing has been used in its strong and weak variants
for constraining the density profiles of clusters. Weak lensing
measures the gravitational tidal field caused by mass distributions,
and thus allows density profiles to be directly inferred. While there
is agreement among most studies of weak cluster lensing that cluster
density profiles are \emph{compatible} with the shape proposed by
\cite{NA96.1,NA97.1} (hereafter NFW), they are typically similarly
well fit by isothermal profiles (e.g.\
\citealt{CL00.1,CL01.1,SH01.1,AT02.1}). The reason is that most of the
weak-lensing signal comes from the cluster regions which surrounded
the scale radius if the clusters had NFW density profiles, and there
the NFW profile has an effective slope close to isothermal.

Strong lensing can happen in the cores of sufficiently dense and
asymmetric clusters and gives rise to highly distorted, arc-like
images. There are now well over 60 clusters known containing arcs with
high length-to-width ratios. Mass models have been constructed for
many of them, but mostly using axially-symmetric or elliptically
distorted mass models with isothermal profiles. Several of these
isothermal models turned out to be spectacularly successful
\citep{KN93.1,KN96.1}. Constructed based on few large arcs, they were
detailed and accurate enough to predict counter-images of arclets
found close to critical curves. \cite{GA03.1} find that the core of
MS~2137 seems to be closer to isothermal, while \cite{KN03.1} give an
example for a cluster which is better fit by NFW than isothermal mass
components.

While models of strong lensing in clusters thus tend to favor density
profiles steeper than expected from numerical simulations,
\cite{SA03.1} followed \cite{MI95.1} in combining the location of
radial and tangential arcs with velocity-dispersion data on the
central cluster galaxies and showed that cluster density profiles
should be substantially less cuspy in their cores than even the NFW
profile. This conclusion hinges on the assumption of axial cluster
symmetry and can be shown to break down for even mildly elliptical
mass models \citep{BA04.1}. However, the situation is obviously
puzzling, and it seems appropriate to ask whether samples of arc
clusters can be successfully modeled with appropriately asymmetric NFW
mass components. This entails two questions; first, can cluster mass
models based on mass components with NFW density profiles be found
which reproduce the observed arcs; and second, are the best-fitting
model parameters within reasonable ranges defined by simulations?

As our sample, we choose clusters which are known to have arcs and
which have been imaged by the Hubble Space Telescope (HST). Our sample
consists of 11 clusters: 3C~220, A~370, Cl~0016, Cl~0024, Cl~0054,
Cl~0939, Cl~1409, Cl~2244, MS~0451, MS~1137, and MS~2137. We model
each cluster with one or more elliptical NFW halos, each of which is
completely defined by its scale convergence, scale radius,
ellipticity, and position angle. In determining the values of these
parameters that best reproduce the observed arcs, we constrain the
mass distribution of the cluster. We find that all 11 clusters can be
successfully modeled using elliptical NFW cluster mass profiles, and
we tightly constrain the parameters defining each cluster's mass
distribution.

The rest of this paper is organized as follows: In \S~\ref{parameter},
we describe our method for estimating the parameters describing a
cluster's set of dark matter halos. In \S~\ref{error}, we define our
error estimation for the derived parameters. In \S~\ref{simulations},
we test our method of estimating lens parameters on a simulated
lensing cluster of known properties.  In \S~\ref{mass}, we outline our
method of calculating the masses of our sample clusters. In
\S~\ref{results}, we present results for each of the 11
clusters. Finally, in \S~7, we discuss our main results and then
summarize the implications of this work. Throughout this paper, we
adopt a spatially flat cosmological model dominated by cold dark
matter and a cosmological constant ($\Omega_\mathrm{m0}=0.3$,
$\Omega_{\Lambda0}=0.7$, $h=0.7$).

\section{Lens Parameter Estimation}
\label{parameter}

Approximately 85\% of the matter in galaxy clusters is dark. If at
all, gas cooling plays a substantial role only in their innermost
cores, where the gas density may be high enough for cooling times to
fall below the Hubble time. It is yet unclear what influence gas
physics may have on strong cluster lensing. While adiabatic gas seems
to have little effect, efficient cooling and star formation may
steepen the density profile very near the cluster center and thus
increase strong-lensing cross sections \citep{PU05.1}. In detail, any
theoretical treatment of baryonic physics on strong cluster lensing
depends on the numerical and artificial viscosity of the gas flow, the
assumed star-formation efficiency, and the combination of a variety of
feedback mechanisms. For simplicity, we shall here model each cluster
in our sample as a combination of purely dark matter halos.

We model each dark matter halo with an asymmetric NFW profile. The
spherical NFW density profile is
\begin{equation}
  \rho(r)=
  \frac{\rho_\mathrm{s}}{(r/r_\mathrm{s})(1+r/r_\mathrm{s})^2}\;,
\end{equation}
where $\rho_\mathrm{s}$ is a characteristic density and $r_\mathrm{s}$
is the scale radius, which describes where the density profile turns
over from $\rho\propto r^{-1}$ to $\rho\propto r^{-3}$.

Following the common thin-lens approximation, the lens is approximated
as a mass sheet perpendicular to the line-of-sight, and the scale
convergence is defined as the ratio of surface mass densities,
$\kappa_\mathrm{s}\equiv\rho_\mathrm{s}r_\mathrm{s}/\Sigma_\mathrm{crit}$,
where $\Sigma_\mathrm{crit}$ is the critical surface mass density,
\begin{equation}
  \Sigma_\mathrm{crit}\equiv\frac{c^2}{4\pi G}\,
  \frac{D_\mathrm{s}}{D_\mathrm{l}D_\mathrm{ls}}\;,
\label{eq:scr}
\end{equation}
with the angular-diameter distances $D_\mathrm{l,s,ls}$ from the
observer to the lens, to the source, and from the lens to the source,
respectively.

Obviously, $\kappa_\mathrm{s}$ is valid for a single source redshift
only. In clusters showing arcs at multiple redshifts,
$\kappa_\mathrm{s}$ needs to be adapted in the following way. Assuming
two source redshifts $z_\mathrm{s}^{(1)}$ and
$z_\mathrm{s}^{(2)}>z_\mathrm{s}^{(1)}$ for simplicity, we refer
$\kappa_\mathrm{s}$ to the lower source redshift
$z_\mathrm{s}^{(1)}$. Fitting the lens model to the data, we adapt
$\kappa_\mathrm{s}$ by the factor
\begin{equation}
  f\equiv\frac{D_\mathrm{ls}^{(2)}}{D_\mathrm{s}^{(2)}}\,
         \frac{D_\mathrm{s}^{(1)}}{D_\mathrm{ls}^{(1)}}
\label{eq:factor}
\end{equation}
for the more distant sources, where $D_\mathrm{s}^{(i)}\equiv
D_\mathrm{s}(z_\mathrm{s}^{(i)})$ and $D_\mathrm{ls}^{(i)}\equiv
D_\mathrm{ls}(z_\mathrm{l}, z_\mathrm{s}^{(i)})$, with $z_\mathrm{l}$
as the lens redshift. Analogous factors are applied for sources at
additional redshifts, if there are any.

To elliptically deform the mass distribution, we alter the potential
to have ellipsoidal rather than axial symmetry. We accomplish this by
defining the surface mass density $\Sigma$ to have a radial dependence
described by the projected elliptical radius,
\begin{equation}
r_\mathrm{e}=\left[(r\cos\theta)^2(1-e)+
(r\sin\theta)^2/(1-e)\right]^{1/2}\;,
\label{eq:re}
\end{equation}
rather than the circular radius $r$. We define the ellipticity
$e=1-b/a$, where $a$ and $b$ are the major and minor axes
respectively, and we define the position angle $\theta$ in degrees
counterclockwise from the $+y$ axis. 

For a given cosmology and halo redshift, an elliptical NFW halo
depends on only four parameters: the scale convergence
$\kappa_\mathrm{s}$, scale radius $r_\mathrm{s}$, ellipticity $e$, and
position angle $\theta$. Numerical simulations predict that the halo
concentration, i.e.~the ratio between the virial radius $r_{200}$ and
the scale radius $r_\mathrm{s}$, is determined by the halo mass,
albeit with considerable scatter
\citep{NA97.1,BU01.1,EK01.1,DO03.2}. This implies that the two
parameters $\rho_\mathrm{s}$ and $r_\mathrm{s}$ characterizing a
spherically symmetric NFW halo are not independent. However, with the
aim of testing numerical results using strong cluster lensing, we do
not adopt any correlation between these two halo parameters.

We identify the arcs on the HST image of a cluster and define them by an
array $(x_i,y_i)$ of $x$ and $y$ positions of the image points
constituting each arc. These points are arranged on a grid with
spacing $\sigma$ in $x$ and $y$. Generally, we take the spacing to be
$\sigma=5$ HST pixels in order to limit the number of arc points. This
array of grid positions forms the data set we use to define a
cluster's arcs.

We use SExtractor \citep{BE96.1} to define the position of each lens
as its center of light on the HST image of the cluster. Then, for a
given set of lens parameters ($\kappa_\mathrm{s}$, $r_\mathrm{s}$,
$e$, $\theta$) for each lens, we use the lensing equations to map the
arc data back to the source plane. This yields a set of points
describing the source. We next use the lensing equations again to map
the source points back to the lens plane by finding all images of all
source points. The result is a set of points that defines the image of
the source reproduced by the lens model. Our goal is to find the
particular values of the lens parameters that yield predicted arcs
that most closely match the arc data, the number of sources predicted
by observations, and reasonably sized sources. Note that our approach
does not require multiple arc-like images of a single source to be
present and identified.

For this purpose we define a $\chi^2$ function of the lens parameters
to quantify the goodness of fit of our model to the data, so that the
minimum of $\chi^2$ produces the best fit. Our $\chi^2$ consists of
three components.

First, we define how well each data point is fit by the image points
by finding the image point that lies closest to a given data point.
We introduce a $\chi^2$ component $\chi_1^2$ that depends on the
distance between each data point and its closest image point. If we
let the $N$ arc data points be ($x_i,y_i$) and the closest predicted
image point to a given data point be
($u_{\mathrm{cl},i},v_{\mathrm{cl},i}$), then this component of the
$\chi^2$ is
\begin{equation}
  \chi_1^2=\frac{1}{N}\,\sum^N_{i=1}\,\left[
    \frac{(x_i-u_{\mathrm{cl},i})^2}{\sigma^2}+
    \frac{(y_i-v_{\mathrm{cl},i})^2}{\sigma^2}
  \right]\;.
\end{equation}
This $\chi_1^2$ thus implicitly assumes that the reproduced image
points are distributed in a Gaussian fashion centered on the data
points, with a standard deviation of $\sigma$.

Second, we define how well each \emph{image} point is predicted by the
\emph{data} points by finding the data point that lies closest to a
given image point. This is not redundant to the $\chi_1^2$ calculation
above because even if every data point has a nearby image point, there
may still be distant, errant image points which are close to no data
points. We thus introduce another $\chi^2$ component $\chi_2^2$ that
depends on the distance between each image point and its closest data
point. If we let the $M$ predicted image points be ($u_j,v_j$) and the
closest data point to a given image point be
($x_{\mathrm{cl},j},y_{\mathrm{cl},j}$), then this component of the
$\chi^2$ is
\begin{equation}
  \chi_2^2=\frac{1}{M}\,\sum^M_{j=1}\,\left[
    \frac{(u_j-x_{\mathrm{cl},j})^2}{\sigma^2}+
    \frac{(v_j-y_{\mathrm{cl},j})^2}{\sigma^2}
  \right]\;,
\end{equation}
assuming again a Gaussian distribution of data with respect to image
points.

Third, we require that arcs which observations indicate are images of
the same source indeed belong to a single source in our model. Some
clusters host several families of arcs, each of which belongs to one
unique source. We require that our model predicts both the number of
sources suggested by observations of a cluster's arcs as well as the
correct correlation between individual arcs and sources, as suggested
by observations. In addition, we require that the predicted source be
small and compact, as lensed sources are commonly observed to be.

Define $N_\mathrm{s}$ as the number of sources suggested by
observations to produce a given cluster's set of arcs. We examine the
$i$th source and its corresponding source points and image points
predicted by our best-fit model to the cluster lens system. If we let
the $P_i$ predicted source points be $(p_{i,j},q_{i,j})$, the mean
$p_i$ position be $\bar p_i$, and the mean $q_i$ position be $\bar
q_i$, then we define the contribution to $\chi^2$ from the source
configuration as
\begin{equation}
  \chi_3^2=\frac{1}{N_\mathrm{s}}\,\sum^{N_\mathrm{s}}_{i=1}\left(
    \frac{1}{P_i}\,\sum^{P_i}_{j=1}\,\left[
      \frac{(p_{i,j}-\bar p_i)^2}{\sigma_\mathrm{s}^2}+
      \frac{(q_{i,j}-\bar q_i)^2}{\sigma_\mathrm{s}^2}
    \right]
  \right)\;,
\end{equation}
where the source points are assumed to have a Gaussian distribution
with standard deviation $\sigma_\mathrm{s}$. The choice of
$\sigma_\mathrm{s}$ is delicate as it controls the relative weight of
the constraints in the lens plane, quantified by $\chi_{1,2}^2$, and
in the source plane, quantified by $\chi_3^2$. Large values of
$\sigma_\mathrm{s}$ may yield best fits with unreasonably large source
configurations, while low values of $\sigma_\mathrm{s}$ may enforce
very small sources at the expense of considerable deviations between
image and data points. Thus, it may be necessary to try several fits
with different choices of $\sigma_\mathrm{s}$ to obtain both good
agreement between image and data points \emph{and} compact sources.

We add these three components to yield our total $\chi^2$,
\begin{equation}
  \chi^2=\chi_1^2+\chi_2^2+\chi_3^2\;.
\end{equation}
By minimizing this quantity we maximize the overlap between data
points and predicted image points, reproduce the number of sources
indicated by observations, and require that the sources be reasonably
small in size. Each combination of ($\kappa_\mathrm{s}$,
$r_\mathrm{s}$, $e$, $\theta$) for each lens describes a different
lens system that produces a different set of source and image points
and hence a different $\chi^2$. We use a downhill-simplex minimization
routine (``amoeba'' from \citealt{PR92.1}) to determine the
combination of each lens's ($\kappa_\mathrm{s}$, $r_\mathrm{s}$, $e$,
$\theta$) that minimizes $\chi^2$. These best-fit values are the ones
we use to define the cluster's lens system.

\section{Error Estimation}
\label{error}

To estimate errors on the cluster dark matter halo parameters derived
according to \S~\ref{parameter}, we employ $\chi^2$ statistics as
outlined in \cite{PR92.1}. We must do so with caution. Formally, the
$\chi^2$ function is the log-likelihood assuming Gaussian
distributions of model points relative to data points. We cannot be
sure that this accurately describes our situation, in which we need to
quantify the deviation between given data and reproduced image
points. Assuming Gaussian likelihood factors, the two contributions
$\chi^2_{1,2}$ quantify the likelihoods of reproducing the data points
with the image points, and of finding image points exclusively near
data points. Taken as another contribution to $\chi^2$, $\chi_3^2$
quantifies the log-likelihood of the sources being well modeled as
Gaussians with widths $\sigma_\mathrm{s}$. Strictly speaking, our
$\chi^2$ is a figure-of-merit function as it quantifies the deviation
of the model from the data in a well-defined sense. We shall, however,
interpret it as a true $\chi^2$ function, i.e.~assuming that all
deviations can be described by appropriate Gaussians.

Recall that for a given lens system, the $\chi^2$ function reaches a
minimum for the best-fit parameter values. Since its gradient vanishes
at the minimum, the $\chi^2$ function can to lowest order of a Taylor
expansion be described as parabolic in the neighborhood of the
minimum. By varying each of the parameters around this minimum, we
thus expect to follow a parabolic section along the parameter axes
through the $\chi^2$ surface.

For a cluster with $N_\mathrm{l}$ lenses, each modeled with an
ellipsoidal NFW density profile, we define a vector of best-fit
parameters $\vec a_\mathrm{bf}=(
\kappa_\mathrm{s,bf,1},r_\mathrm{s,bf,1},
e_\mathrm{bf,1},\theta_\mathrm{bf,1},\ldots,
\kappa_\mathrm{s,bf,N_\mathrm{l}},r_\mathrm{s,bf,N_\mathrm{l}},
e_\mathrm{bf,N_\mathrm{l}},\theta_\mathrm{bf,N_\mathrm{l}})$. We then
calculate the $16\,N_\mathrm{l}^2$ components of the curvature matrix
$\alpha_{kl}$ by noting that
\begin{equation}
  \alpha_{kl}=\frac{1}{2}\,\left.
    \frac{\partial^2\chi^2}{\partial a_k\partial a_l}
  \right|_{\vec a=\vec a_\mathrm{bf}}\;,
\end{equation}
where $k$ and $l$ vary from 1 to $4N_\mathrm{l}$. The partial
derivatives can be calculated from the parabolic fits along all
parameter axes to the local $\chi^2$ function. The covariance matrix
is the inverse of the curvature matrix, $C=\alpha^{-1}$, and the
1-$\sigma$ error in each parameter is $\delta a_k=\sqrt{C_{kk}}$. We
generally find these errors to be of order a few percent. An
interesting approach for estimating errors on strong-lensing model
parameters based on Monte-Carlo Markov Chains was recently proposed by
\cite{BR05.3}.

\section{Comparison of Parameter Estimation Method with Simulations}
\label{simulations}

To test our method of lens parameter estimation outlined in
\S~\ref{parameter}, we use a numerically simulated galaxy cluster
producing arcs.

The cluster was kindly made available by Klaus Dolag. It was obtained
by re-simulating at higher resolution a patch of a pre-existing large
scale numerical simulation of the $\Lambda$CDM model with parameters
$\Omega_\mathrm{m0}=0.3$, $\Omega_{\Lambda0}=0.7$,
$H_0=70\,\mathrm{km\,s^{-1}\,Mpc^{-1}}$ and normalization
$\sigma_8=0.9$. The ``ZIC'' technique used is described in detail in
\citet{TO97.2}. The cluster has redshift $z=0.3$ and a virial mass of
$M_\mathrm{vir}=2.29\times10^{15}\,h^{-1}\,M_\odot$. The particle mass
in the re-simulation is
$m_\mathrm{part}=1.3\times10^9\,h^{-1}\,M_\odot$. The gravitational
softening is set to $5\,h^{-1}\,\mathrm{kpc}$.

\begin{figure}[!t]
\plotone{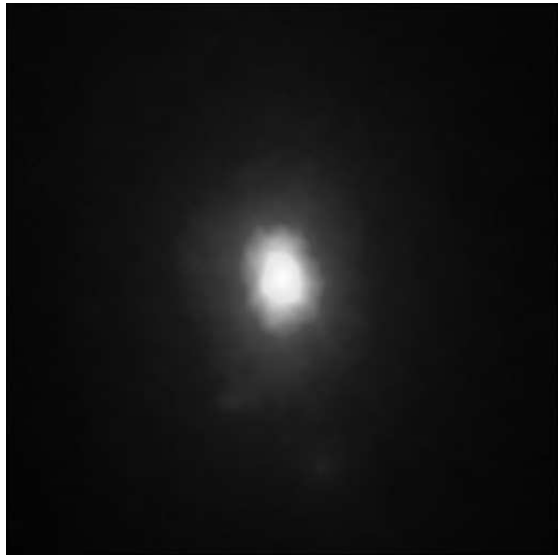}
\caption{A $\sim 370''\times370''$ surface density map of the
numerical cluster used to test our lens parameter estimation
method. \label{fig:kmap}} 
\end{figure}

We simulate lensing by this massive cluster using standard ray-tracing
techniques. First, the particles contained in a cube of
$3\,h^{-1}\,\mathrm{Mpc}$ comoving side length are selected. Then, to
produce a two-dimensional density field, their masses are projected
along the line of sight, interpolating their positions onto a regular
grid of $256\times256$ pixels using the \emph{Triangular Shaped Cloud}
method \citep{HO88.1}. This surface density map, shown in
Figure~\ref{fig:kmap}, is used as the lens plane in the following
lensing simulation.

\begin{figure}[!t]
\plotone{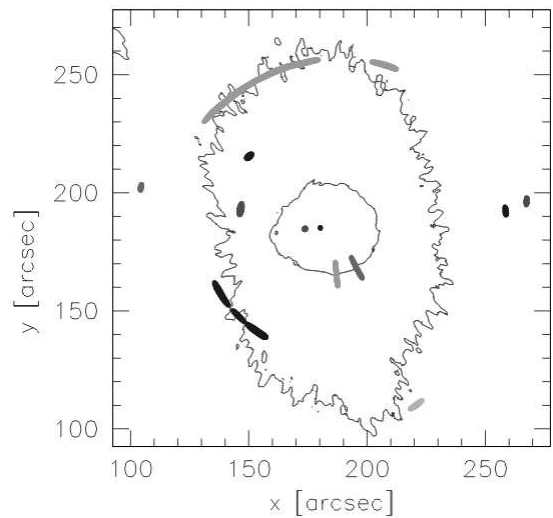}
\caption{The central region of Figure~\ref{fig:kmap}, illustrating the
critical lines for the numerical cluster as well as examples of arcs
produced by this cluster lens. We used such systems of arcs to test
our method of lens parameter estimation. \label{fig:critarcs}}
\end{figure}

A bundle of $2048\times2048$ light rays is traced from the observer
through the central quarter of the lens plane and their deflection due
to the cluster mass distribution is calculated as described in several
earlier papers (see e.g.\ \citealt{ME00.1,ME01.1,ME03.1,ME03.2}). The
arrival positions of the light rays on the source plane, which we
place at redshift $z_\mathrm{s}=2$, are used to reconstruct the lensed
images of several sources distributed around the caustic curves so as
to produce strong lensing features. The sources are modeled as
ellipses, with random orientation and axial ratios randomly drawn with
equal probability from $[0.5,1]$. Their equivalent diameter (the
diameter of the circle enclosing the same area as the source) is
$r_\mathrm{e}=1''$. Several arc configurations have been used to test
our method, and some of these are illustrated in
Figure~\ref{fig:critarcs}.

We conduct a blind test of our parameter estimation method by applying
it to the simulated cluster and arcs without knowledge of any of the
cluster's physical properties but its position. We permit knowledge of
the cluster's position because when we model an HST cluster, we
determine the positions of its galaxy lenses on the HST image with
SExtractor.

We model the simulated cluster based on the cluster position and arc
points, which is the same information we have when we model HST
clusters. Applying the parameter estimation method described in
\S~\ref{parameter}, we estimate the scale convergence, scale radius,
ellipticity, and position angle of the simulated cluster. We then
compare with the true values of these parameters in the simulated
cluster mass distribution.

\begin{deluxetable*}{llllll}
\tablewidth{0pt}
\tablecolumns{6}
\tablecaption{Best-fit parameters to lenses in the cluster
  sample.}
\tablehead
  {\colhead{Cluster} &
   \colhead{Lens} &
   \colhead{$\kappa_\mathrm{s}$} &
   \colhead{$r_\mathrm{s}\,(h^{-1}\,\mathrm{kpc})$} &
   \colhead{$e$} &
   \colhead{$\theta\,(^\circ)$}}
\startdata
Cl~2244$-$02 & &
$0.178\pm0.003$ & $260\pm20$ & $0.113\pm0.005$ & $179\pm1$ \\
Abell~370 & G1 &
$0.164\pm0.007$\tablenotemark{a} & $254\pm2$ &
$0.28\pm0.01$ & $78\pm2$ \\
& G2 &
$0.165\pm0.004$\tablenotemark{b} & $212\pm1$ &
$0.073\pm0.004$ & $167\pm1$ \\ 
3C~220.1 & &
$0.178\pm0.002$ & $226\pm4$ & $0.265\pm0.005$ & $25\pm1$ \\
MS~2137$-$23 & &
$0.67\pm0.02$ & $64\pm2$ & $0.11\pm0.02$ & $95\pm3$ \\
Cl~140933$+$5226 & &
$0.099\pm0.001$ & $241\pm2$ & $0.033\pm0.001$ & $79\pm2$ \\
MS~0451.6$-$0305 & &
$0.276\pm0.006$\tablenotemark{c} &
$262\pm8$ & $0.215\pm0.004$ & $156\pm1$ \\
MS~1137$+$66 & &
$0.256\pm0.002$ & $279\pm1$ & $0.143\pm0.005$ & $58\pm2$ \\
Cl~0054$-$27 & G1 &
$0.047\pm0.001$ & $340\pm20$ & $0.175\pm0.003$ & $92\pm1$ \\
& G2 &
$0.100\pm0.001$ & $259\pm7$ & $0.14\pm0.01$ & $18.6\pm0.8$ \\
Cl~0016$+$1609 & DG 256 &
$0.102\pm0.003$ & $270\pm10$ & $0.123\pm0.002$ & $64.7\pm0.4$ \\
& DG 251 &
$0.087\pm0.001$ & $192\pm4$ & $0.121\pm0.005$ & $11.0\pm0.5$ \\
& DG 224 &
$0.219\pm0.004$ & $261\pm7$ & $0.1691\pm0.0008$ & $7.1\pm0.4$ \\
Cl~0939$+$4713 & G1 &
$0.126\pm0.005$ & $136\pm6$ & $0.364\pm0.008$ & $2.65\pm0.03$ \\
& G2 &
$0.114\pm0.002$ & $190\pm10$ & $0.215\pm0.008$ & $73\pm1$ \\
& G3 &
$0.156\pm0.005$ & $170\pm1$ & $0.044\pm0.003$ & $87\pm3$ \\
Cl~0024$+$17 & \#362 &
$0.158\pm0.003$ & $198\pm2$ & $0.059\pm0.002$ & $6.6\pm0.2$ \\
& \#374 &
$0.170\pm0.002$ & $250\pm7$ & $0.153\pm0.006$ & $135\pm7$ \\
& \#380 &
$0.116\pm0.003$ & $285\pm2$ & $0.0020\pm0.0001$ & $58.1\pm0.8$ \\
\enddata
\tablenotetext{a}{The value of $\kappa_\mathrm{s}$ in the source plane
  of the giant arc A0. In the source plane of the arc pair B2/B3,
  $\kappa_\mathrm{s}=0.181\pm0.008$ and in the source plane of the
  radial arc R, $\kappa_\mathrm{s}=0.24\pm0.01$.}
\tablenotetext{b}{The value of $\kappa_\mathrm{s}$ in the source plane
  of the giant arc A0. In the source plane of the arc pair B2/B3,
  $\kappa_\mathrm{s}=0.183 \pm 0.004$ and in the source plane of the
  radial arc R, $\kappa_\mathrm{s}=0.242 \pm 0.006$.}
\tablenotetext{c}{The value of $\kappa_\mathrm{s}$ in the source plane
  of the upper arc, ARC2. In the source plane of the lower arc, ARC1,
  $\kappa_\mathrm{s}=0.55\pm0.01$.}
\label{tbl-1}
\end{deluxetable*}

\begin{deluxetable*}{llllll}
\tablewidth{0pt}
\tablecolumns{6}
\tablecaption{Estimated cluster lens masses. \label{tbl-2}}
\tablehead
  {\colhead{Cluster} &
   \colhead{Lens} &
   \colhead{$z_\mathrm{arc}$} &
   \colhead{$r_\mathrm{s}$} &
   \colhead{$M\,(\le r_\mathrm{s})$} &
   \colhead{Reference} \\ &&&
   \colhead{$(h^{-1}\,\mathrm{kpc})$} &
   \colhead{$(h^{-1}\,M_\odot)$}&}
\startdata
Cl~2244$-$02 & &
2.237 & 260 & $9.33\times10^{13}$ & 1 \\
Abell~370 & G1 &
0.724/0.806/1.3 & 254 & $1.31\times10^{14}$ & 2 \\
& G2 &
0.724/0.806/1.3 & 212 & $9.21\times10^{13}$ & 2 \\
3C~220.1 & &
1.49  & 226 & $7.65\times10^{13}$ & 3 \\
MS~2137$-$23 & &
1.501 &  64 & $2.3 \times10^{13}$ & 4 \\
Cl~140933$+$5226 & &
2.811 & 241 & $3.78\times10^{13}$ & 5 \\
MS~0451.6$-$0305 & &
0.917/2.911 & 262 & $2.38\times10^{14}$ & 6 \\
Cl~0939$+$4713 & G1 &
3.98 & 136 & $1.52\times10^{13}$ & 7 \\
& G2 &
3.98 & 190 & $2.68\times10^{13}$ & 7 \\
& G3 &
3.98 & 170 & $2.92\times10^{13}$ & 7 \\
Cl~0024$+$17 & \#362 &
1.675 & 198 & $4.75\times10^{13}$ & 8 \\
& \#374 &
1.675 & 250 & $8.17\times10^{13}$ & 8 \\
& \#380 &
1.675 & 285 & $7.21\times10^{13}$ & 8 \\
\enddata

\tablerefs{(1) \cite{SM97.2}; (2) \cite{BE99.1}; (3) \cite{OT00.1};
(4) \cite{GA05.1}; (5) \cite{LU00.1}; (6) \cite{BO04.2};
(7) \cite{TR97.1}; (8) \cite{BR00.1}.}

\end{deluxetable*}

In the simulation, the cluster is found to be well-described as an NFW
ellipsoid with $\kappa_\mathrm{s}=0.54$, $r_\mathrm{s}=92.4''$,
$e=0.18$, and $\theta=99^\circ$. With our arc modeling method, we
found the cluster's best-fit parameters to be
$\kappa_\mathrm{s}=0.55$, $r_\mathrm{s}=88.9''$, $e=0.17$, and
$\theta=96^\circ$, which are within 4\% of the parameter values used
to approximate the cluster. This is a convincing match, even more so
because the true cluster parameters are the results of a fit and carry
errors themselves. Hence, we proceed with confidence in our lens
parameter estimation routine to model our sample of HST clusters.

\section{Cluster Mass Estimates}
\label{mass}

If we know the lens parameters of a cluster and the source and lens
redshifts, we can determine the mass of the lens within a given
radius. We assume the lens is described by an NFW density profile,
where the characteristic density is the density at the scale radius,
given by $\rho_\mathrm{s}=\kappa_\mathrm{s}
\Sigma_\mathrm{crit}/r_\mathrm{s}$. If we define $x\equiv
r/r_\mathrm{s}$, then the mass contained within a (three-dimensional)
radius $R$ is
\begin{eqnarray}
  M(\le R)&=&4\pi r_\mathrm{s}^3\rho_\mathrm{s}\,
  \int_0^R\frac{x^2\d x}{x(1+x)^2}\nonumber\\&=&
  4\pi\Sigma_\mathrm{crit}\kappa_\mathrm{s}r_\mathrm{s}^2\,
  \left[\ln(1+y)-\frac{y}{1+y}\right]\;,
\end{eqnarray}
where $y\equiv R/r_\mathrm{s}$. The critical surface-mass density is
given in equation~(\ref{eq:scr}).

\begin{figure*}[!t]
\centering
\epsscale{0.85}
\plotone{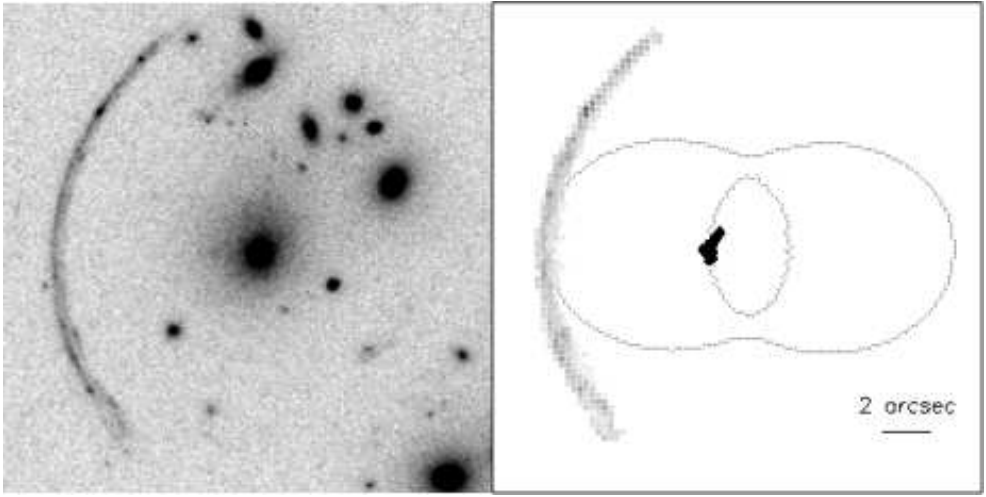}
\caption{A $20''\times20''$ section of the F555W WFPC2 image of
  Cl~2244$-$02 (left) and the same section of the best fit
  (right). The plot on the right illustrates the reproduced giant arc
  (gray points), the predicted source (black points), and the lens's
  critical curves. The gray image points are scaled such that darker
  grays denote brighter points in the image. \label{fig:cl2244}}
\end{figure*}

\begin{figure*}[t]
\centering
\epsscale{0.85}
\plotone{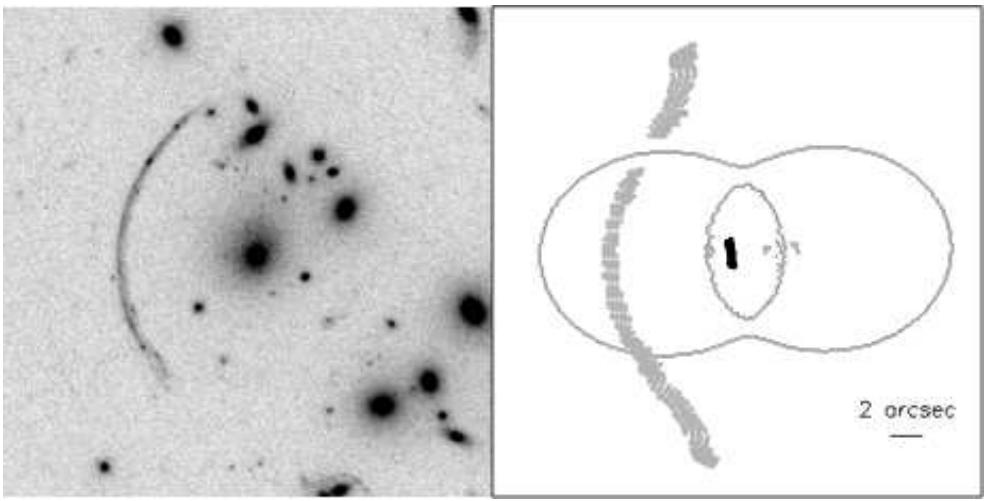}
\caption{As Figure~\ref{fig:cl2244}, but enlarged to $30''\times30''$
  and using a model with lens parameters 3-$\sigma$ greater than the
  best-fit parameters (right). \label{fig:cl2244sig3}}
\end{figure*}

For each lens in a cluster, we calculate the projected mass contained
within the lens's scale radius. We note that for the clusters in our
sample without published arc redshifts (MS~1137, Cl~0054, and
Cl~0016), we cannot estimate the lens masses.

\section{Parameterizations and Mass Estimates of the Clusters}
\label{results}

Here we present the best fit to each cluster based on the observed
arcs, as described in \S~\ref{parameter}. We will discuss each cluster
individually. For a summary of the best-fit parameters for all sampled
clusters, see Table~\ref{tbl-1}. Our mass results are summarized in
Table~\ref{tbl-2}, which includes the reference for each cluster's 
arc redshifts.

\begin{itemize}

\item \textbf{Cl~2244$-$02}: The cluster Cl~2244$-$02 has a redshift
  $z=0.33$ and hosts a spectacular tangential arc
  \citep{SM97.2}. \citet{LY89.1} discovered this giant luminous arc,
  which is located near the cluster center. The arc is a partial
  Einstein ring, and we take the lens to be the large galaxy seen in
  Figure~\ref{fig:cl2244} near where the Einstein ring is centered.

  Although the errors we calculated for the best-fit parameters are as
  small as half a percent, they are indeed realistic. In
  Figure~\ref{fig:cl2244sig3} we illustrate what the predicted source
  and images would be for a lens defined by parameters 3-$\sigma$
  greater than our best-fit values to Cl~2244$-$02. The predicted
  giant arc is much larger than that observed, and is broken into two
  sections. Also, two small images are produced near the source that
  are not seen in observations.

  Changing the lens parameters by only 3-$\sigma$ significantly alters
  the lensed images, which indicates that the lens parameters must be
  tightly constrained around our best-fit values. We use this example
  to justify the small errors we find for Cl~2244$-$02, and extend the
  argument also to the small errors we find for clusters that follow.

\begin{figure*}[!t]
\centering
\epsscale{0.85}
\plotone{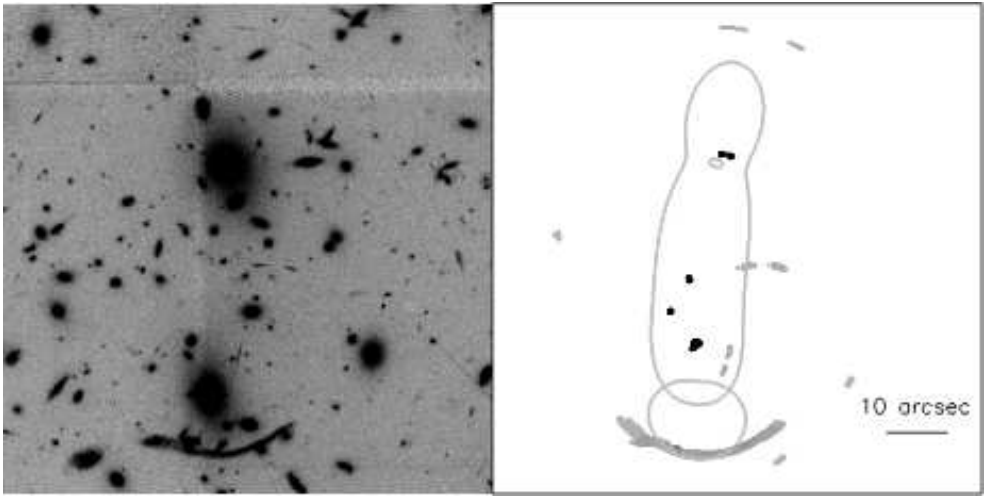}
\caption{As Figure~\ref{fig:cl2244}, showing an $80''\times80''$ section
  of the F675W WFPC2 image of Abell~370. \label{fig:a370}}
\end{figure*}

\begin{figure*}[!t]
\centering
\epsscale{0.85}
\plotone{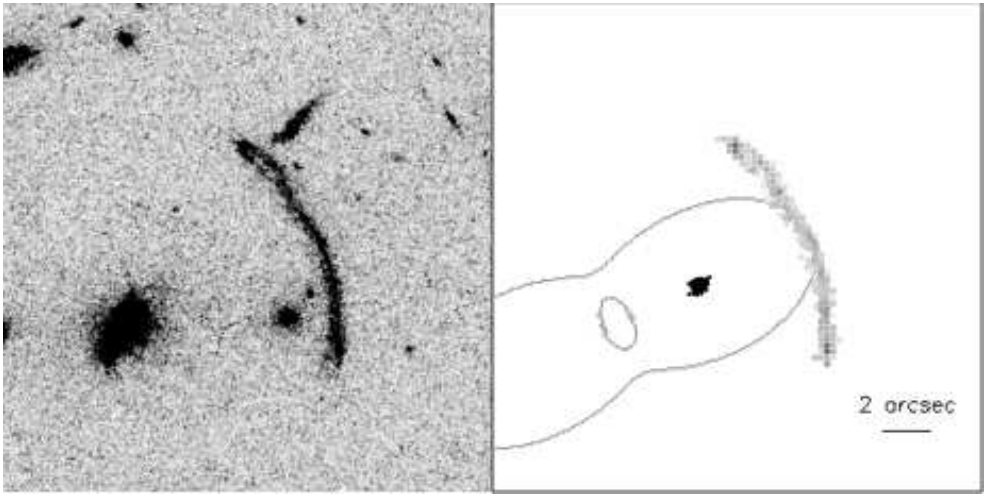}
\caption{As Figure~\ref{fig:cl2244}, showing a $20''\times20''$ section
  of the F555W WFPC2 image of 3C~220.1. \label{fig:3c220}}
\end{figure*}

\item \textbf{Abell~370}: The rich cluster Abell~370 is at redshift
  $z=0.375$ \citep{AB98.1} and has a bimodal mass distribution with
  two cD galaxies. These two galaxies mark the centers of two mass
  components in our model, and we identify the northern cD as G1 and
  the southern cD as G2.

  By the classification of \cite{BE99.1}, the giant arc is A0, the
  nearby radial and two tangential arcs are R, B2, and B3, and the
  upper tangential arcs are A1 and A2. These six arcs, as well as the
  cD galaxy lenses, are visible in Figure~\ref{fig:a370}.

  We assume A1 and A2 to be at the same redshift as the dominant arc
  A0. They are likely to form a double image of the same source. Arcs
  B2 and B3 share the same redshift and are images of a separate
  source; R is an image of yet another source
  \citep{BE99.1}. Accordingly, our best-fit model produces four
  independent sources for these six arcs.

  Our best fit also produces three additional images that we did not
  identify in our initial arc sample. They correspond to fuzzy patches
  on the HST image which may reflect actual images. The exact number
  of arcs in the cluster is unclear; \cite{BE99.1} identify as many as
  81 arclets. Further study would offer more insight into the complex
  nature of Abell~370.

\begin{figure*}[!t]
\centering
\epsscale{0.85}
\plotone{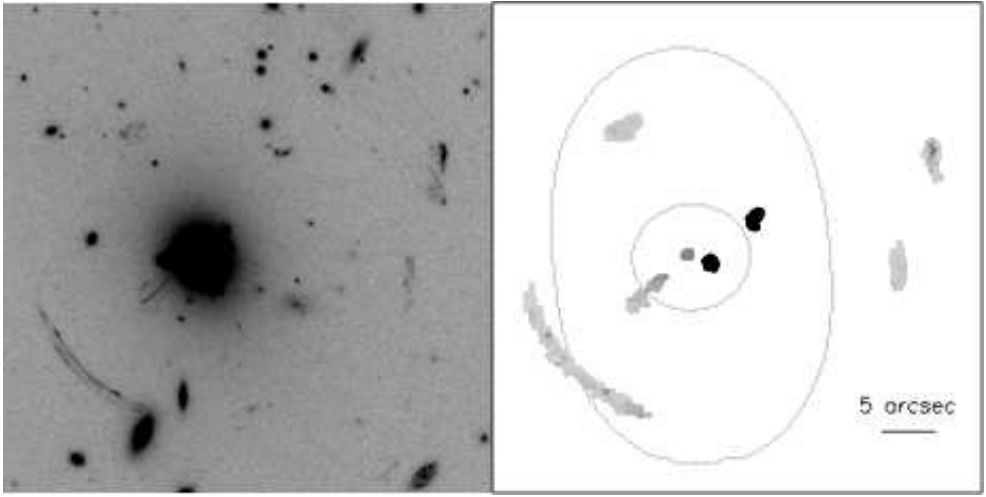}
\caption{As Figure~\ref{fig:cl2244}, showing a $45''\times45''$ section
  of the F702W WFPC2 image of MS~2137$-$23. \label{fig:ms2137}}
\end{figure*}

\begin{figure*}[!t]
\centering
\epsscale{0.85}
\plotone{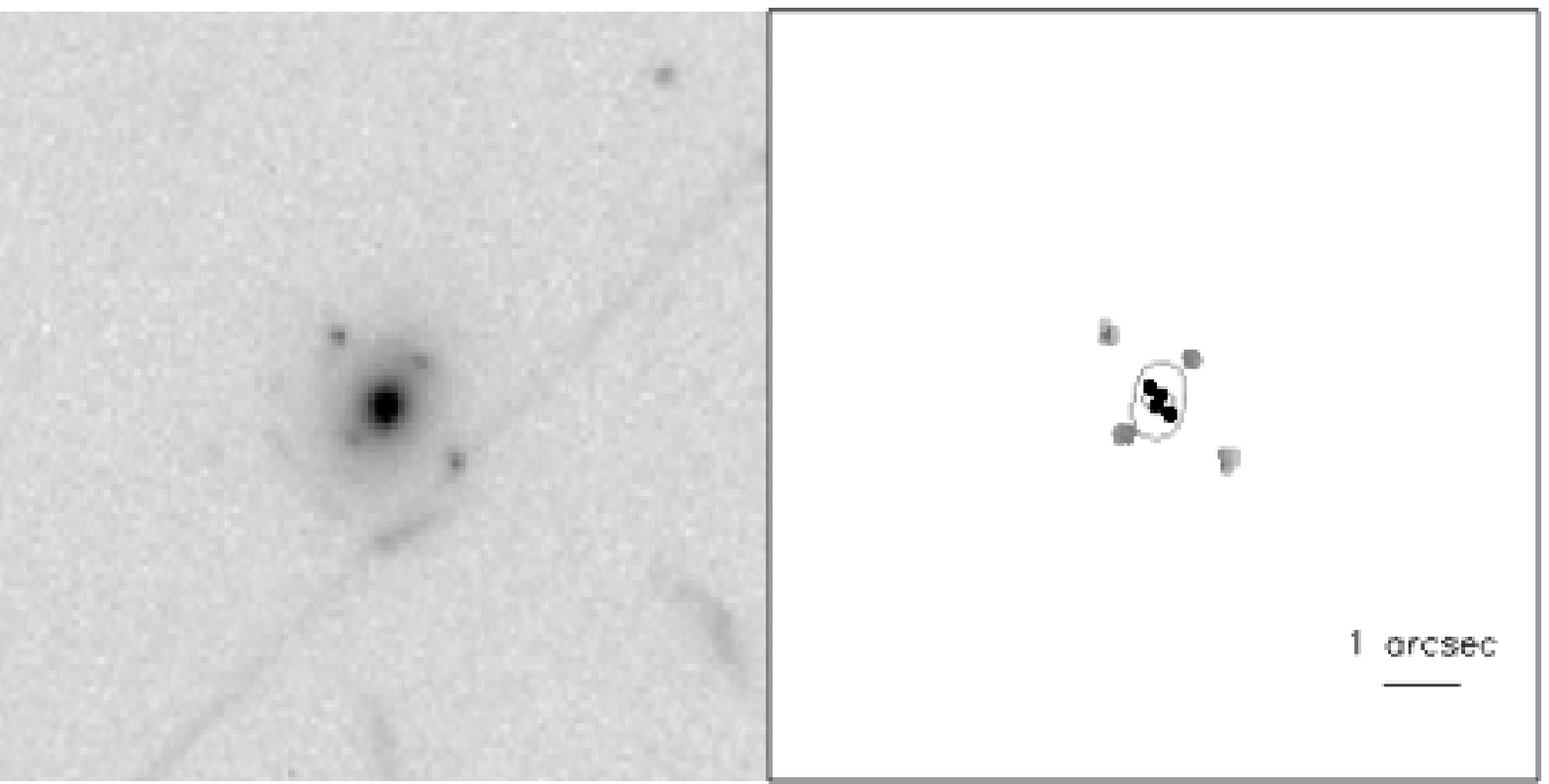}
\caption{As Figure~\ref{fig:cl2244}, showing a $10''\times10''$ section
  of the F702W WFPC2 image of Cl~140933$+$5226. \label{fig:cl1409}}
\end{figure*}

\item \textbf{3C~220.1}: The cluster containing the radio galaxy
  3C~220.1 is at redshift $z=0.62$ and hosts a giant arc
  \citep{OT00.1}. The lens responsible for producing this arc is
  3C~220.1, which is identifiable as the largest source in the HST
  image on the left of Figure~\ref{fig:3c220}.

\item \textbf{MS~2137$-$23}: MS~2137$-$23 is a rich cluster at
  redshift $z=0.313$ that contains a giant luminous arc and several
  arclets \citep{GA02.2}. The cluster is dominated by a single cD
  galaxy, which we define as the lens. Adopting the classification
  system of \cite{GA02.2}, the giant arc is A0, the radial arc near
  the cD galaxy is A1, the arc to the left of the cD galaxy in
  Figure~\ref{fig:ms2137} is A2, and the two arcs to the right of the cD
  galaxy are A4 and A5. The lens and the five arcs are visible in the
  HST image of MS~2137$-$23 shown in Figure~\ref{fig:ms2137}.

  \cite{GA05.1} also fit an NFW model to MS~2137$-$23, and the
  ellipticity found in that paper is within the error bars of our
  best-fit ellipticity. However, our scale radius is $\sim30\%$
  smaller than, and our scale convergence is twice as large as, the
  corresponding values found in \cite{GA05.1}. Despite these
  discrepancies, both our findings and those of \cite{GA05.1} agree
  that the concentration of the cD galaxy in MS~2137$-$23 is high.

  Our lens model predicts an additional arc, near the center of the cD
  galaxy, which was also discussed and possibly detected by
  \cite{GA03.1}. \cite{GA02.2} suggests that the arcs in MS~2137$-$23
  can be categorized into two different systems, each with a unique
  source. The arcs A0, A2, and A4 are images of one source, and A1 and
  A5 are images of a separate source. We use these correlations to
  constrain the lens parameters, and Figure~\ref{fig:ms2137} illustrates
  the two unique sources predicted by our best-fit lens model.

\begin{figure*}[!t]
\centering
\epsscale{0.85}
\plotone{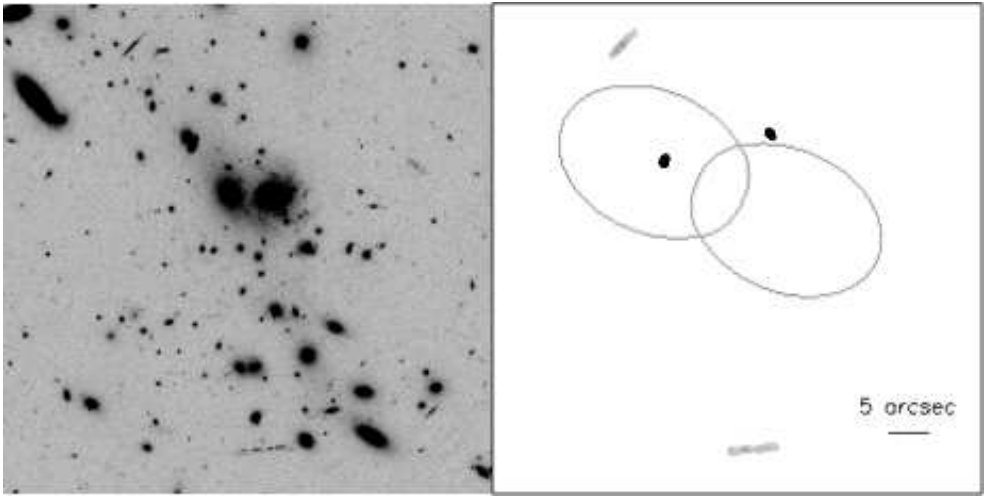}
\caption{As Figure~\ref{fig:cl2244}, showing a $60''\times60''$ section
  of the F702W WFPC2 image of MS~0451.6$-$0305. \label{fig:ms0451}}
\end{figure*}

\begin{figure*}[t]
\centering
\epsscale{0.85}
\plotone{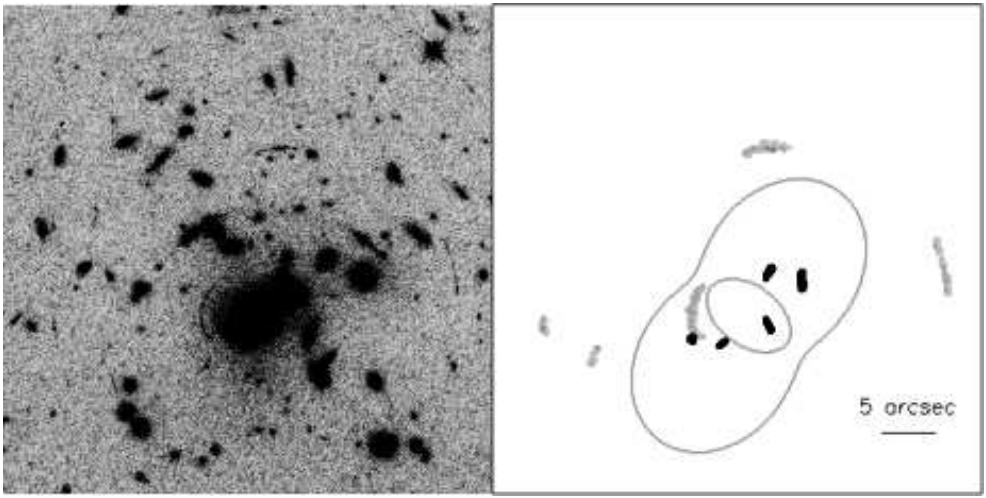}
\caption{As Figure~\ref{fig:cl2244}, showing a $45''\times45''$ section
  of the F814W WFPC2 image of MS~1137$+$66. \label{fig:ms1137}}
\end{figure*}

\item \textbf{Cl~140933$+$5226}: \cite{FI98.1} discovered a quadruple
  lens in the $z=0.46$ cluster Cl~140933$+$5226, also known as
  3C~295. The Einstein cross, as well as the lens galaxy at its
  center, is visible in Figure~\ref{fig:cl1409}.

  As expected, the four images are due to a single source at the
  center of the Einstein cross.

\item \textbf{MS~0451.6$-$0305}: The cluster MS~0451.6$-$0305 is
  located at $z=0.55$ and hosts two tangential arcs
  \citep{BO04.2}. The two arcs, as well as the cD galaxy lens, are
  apparent in the HST image in Figure~\ref{fig:ms0451}.  Following the
  classification of \cite{BO04.2}, the lower arc in the figure is ARC1
  and the upper arc is ARC2.

  \cite{BO04.2} note that the two arcs in MS~0451.6$-$0305 are at
  different redshifts and thus are not images of a single
  source. These two separate sources can be seen in our fit, shown in
  Figure~\ref{fig:ms0451}.

\begin{figure*}[t]
\centering
\epsscale{0.85}
\plotone{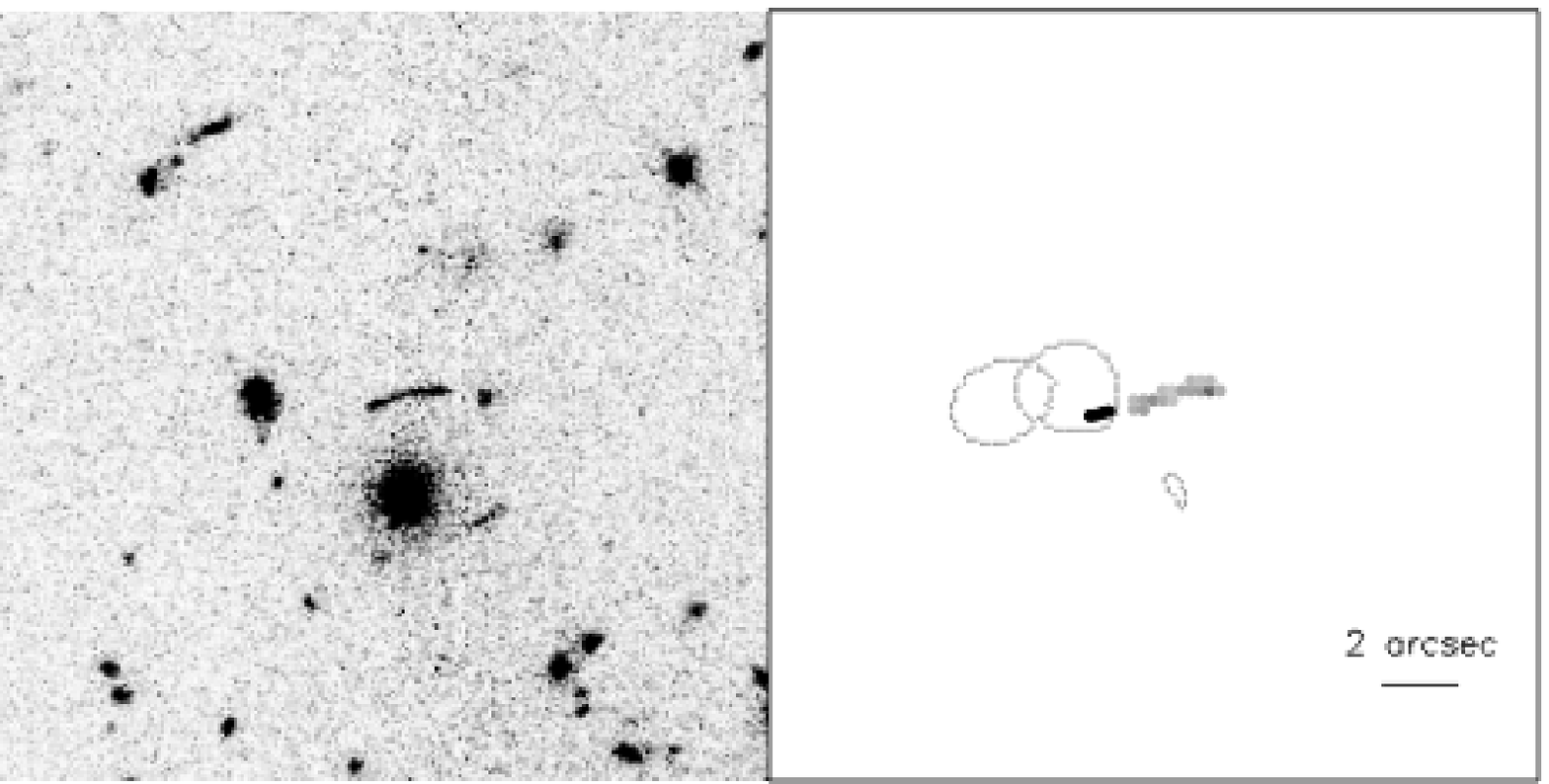}
\caption{As Figure~\ref{fig:cl2244}, showing a $20''\times20''$ section
  of the F555W WFPC2 image of Cl~0054$-$27. \label{fig:cl0054}}
\end{figure*}

\begin{figure*}[t]
\centering
\epsscale{0.85}
\plotone{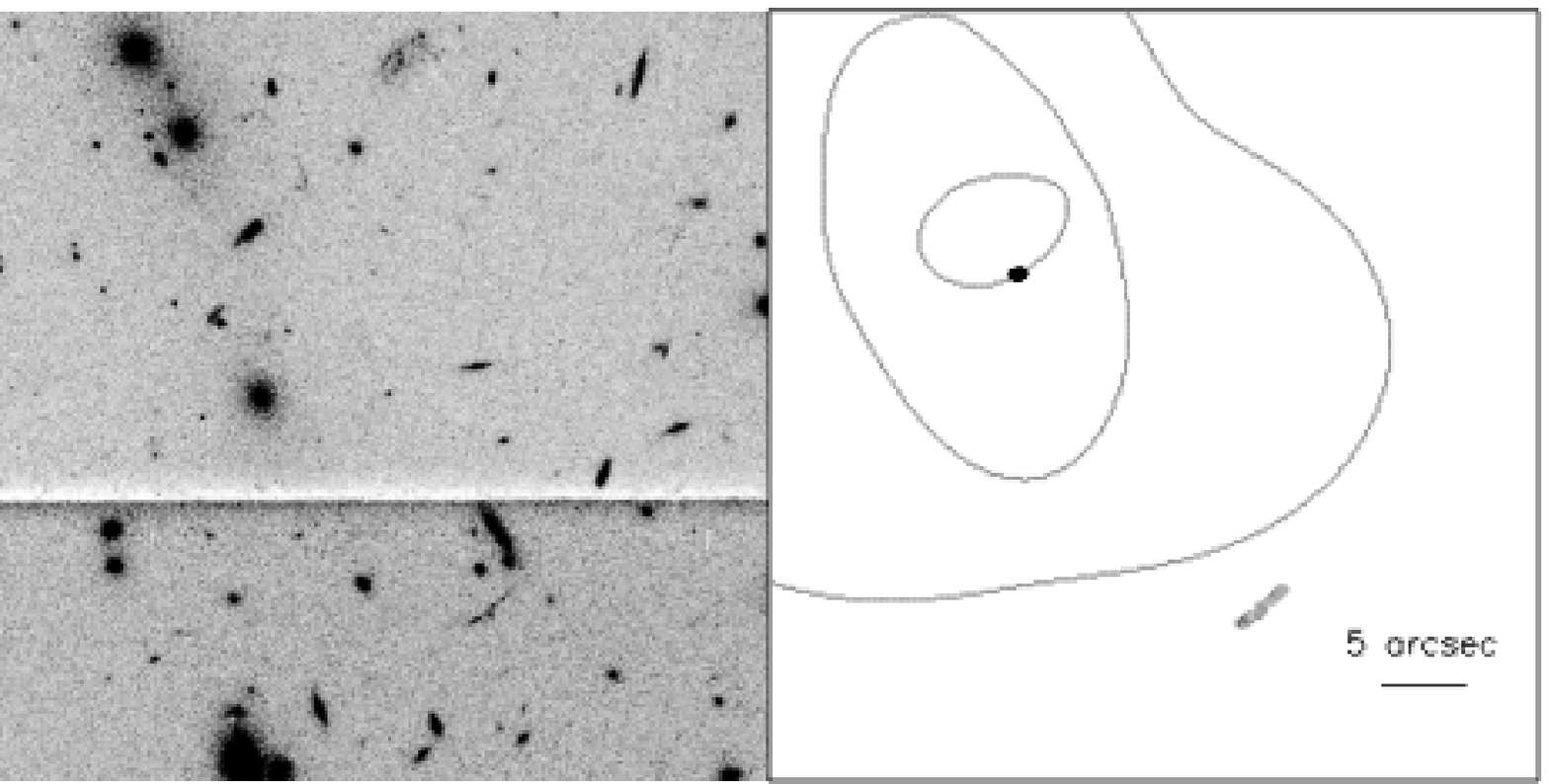}
\caption{As Figure~\ref{fig:cl2244}, showing a $45''\times45''$ section
  of the F555W WFPC2 image of Cl~0016$+$1609. \label{fig:cl0016}}
\end{figure*}

\item \textbf{MS~1137$+$66}: The lensing cluster MS~1137$+$66 at
  $z=0.783$ \citep{CL98.1} has the highest redshift of the clusters in
  our sample and hosts several faint arcs. We model five arcs, and
  because there are no published redshifts for the arcs, for
  simplicity we assume that they are all at the same redshift. The
  lens in our model is the cD galaxy at the center of the cluster.

  The lens defined by the parameters in Table~\ref{tbl-1} accurately
  reproduces the five observed arcs, as seen in
  Figure~\ref{fig:ms1137}. Each arc has its own independent source.

  We cannot estimate the mass of the cD galaxy lens because there are
  no published redshifts for the arcs in MS~1137$+$66.

\begin{figure*}[t]
\centering
\epsscale{0.85}
\plotone{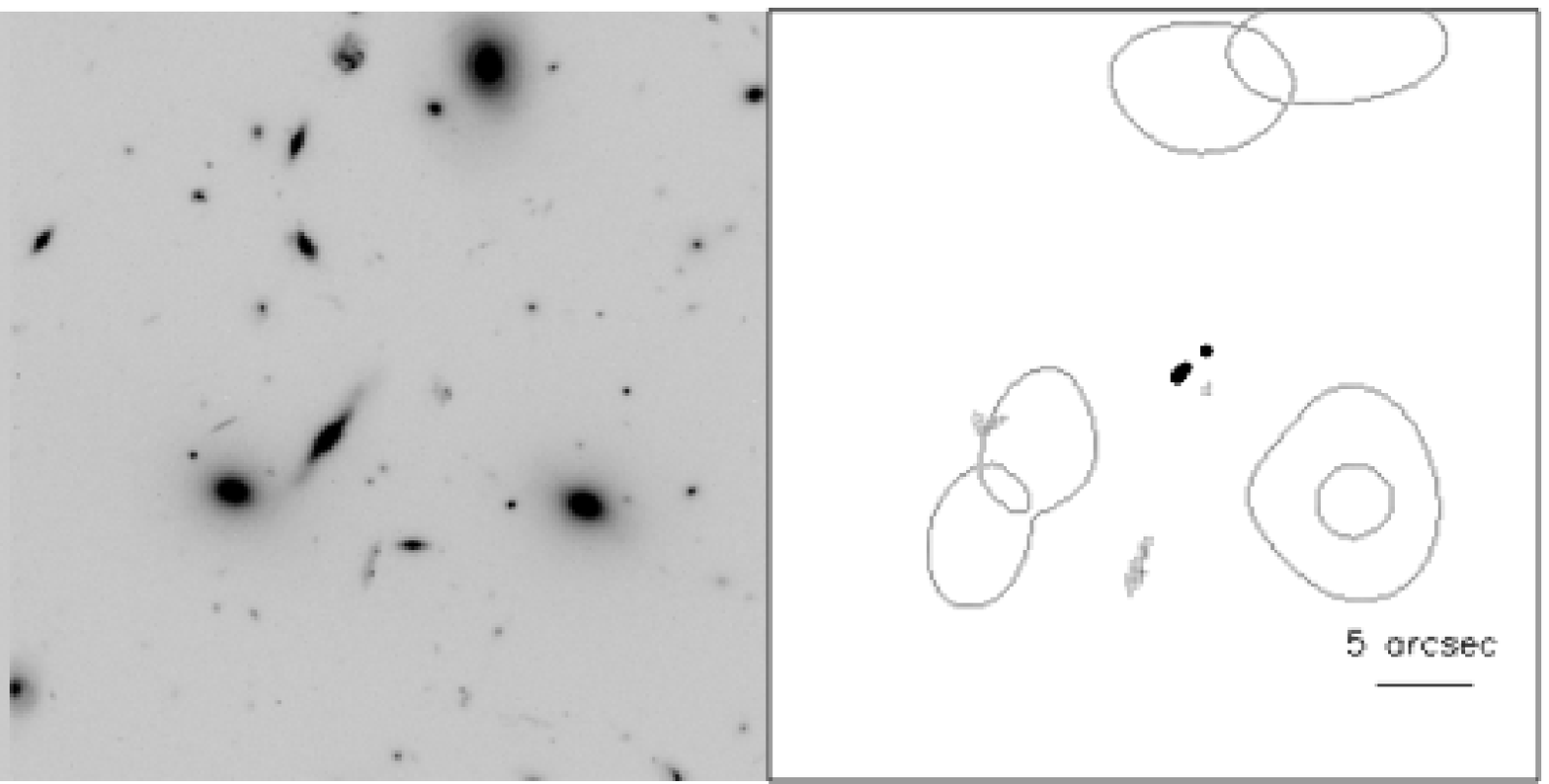}
\caption{As Figure~\ref{fig:cl2244}, showing a $40''\times40''$ section
  of the F702W WFPC2 image of Cl~0939$+$4713. \label{fig:cl0939}}
\end{figure*}

\begin{figure*}[t]
\centering
\epsscale{0.85}
\plotone{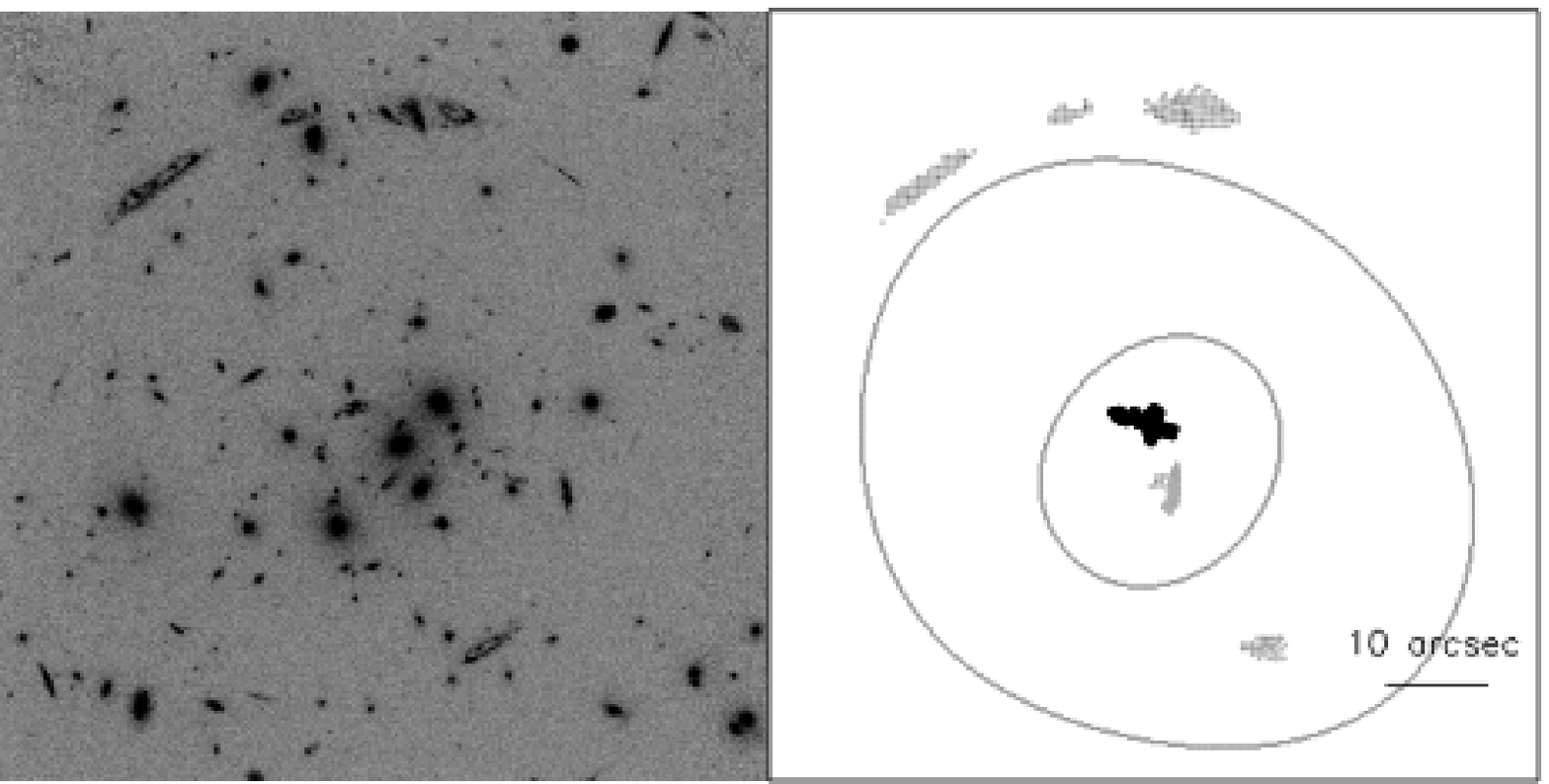}
\caption{As Figure~\ref{fig:cl2244}, showing a $75''\times75''$ section
  of the F450W WFPC2 image of Cl~0024$+$17. \label{fig:cl0024}}
\end{figure*}

\item \textbf{Cl~0054$-$27}: The cluster Cl~0054$-$27 is located at
  redshift $z=0.56$ and has one lensed arc \citep{SM97.3}. To
  reproduce the arc, we take the two lenses in the cluster to be the
  central galaxy (which we call G1) and the upper left galaxy (which
  we call G2) in Figure~\ref{fig:cl0054}.

  These two lenses combined reproduce the observed arc accurately. A
  mass estimate of the two lenses in Cl~0054$-$27 is impossible
  because there is no published redshift for the arc.

\item \textbf{Cl~0016$+$1609}: The cluster Cl~0016$+$1609, at redshift
  $z=0.545$, has one thin lensed arc \citep{LA96.1}. The three
  approximately collinear elliptical galaxies seen in
  Figure~\ref{fig:cl0016} define the center of the cluster and are the
  lenses we use to reproduce the observed arc. These three giant
  galaxies are, from top to bottom in Figure~\ref{fig:cl0016}, DG~256,
  DG~251, and DG~224.

  The combination of these three lenses reproduces the single thin arc
  convincingly. Because there is no published redshift for the arc, we
  are unable to estimate the masses of the lenses in Cl~0016$+$1609.

\item \textbf{Cl~0939$+$4713}: The cluster Cl~0939$+$4713 is at
  redshift $z=0.41$ and has three radial arcs near its center
  \citep{SE96.2}. Three giant elliptical galaxies are visible in
  Figure~\ref{fig:cl0939}, and these galaxies make up the cluster's
  core. We take these galaxies to be the gravitational lenses, and
  call the upper one in Figure~\ref{fig:cl0939} G1, the leftmost one G2,
  and the rightmost one G3. The three arcs are also visible in
  Figure~\ref{fig:cl0939}.

  Defined by their best-fit parameters given in Table~\ref{tbl-1}, the
  three lenses are able to reproduce the three arcs observed in
  Cl~0939$+$4713. \cite{TR97.1} argue that two of the arcs in
  Cl~0939$+$4713 are likely images of the same source, whereas the
  third arc is probably an image of a separate galaxy. Consistent with
  this expectation, our model predicts two unique sources for the arc
  system.

\item \textbf{Cl~0024$+$17}: The cluster Cl~0024$+$17 has redshift
  $z=0.395$ and hosts five images of a single background galaxy
  \citep{OT04.1}. The three central elliptical galaxies in the
  cluster, seen roughly collinear in Figure~\ref{fig:cl0024}, act as the
  lenses for this system. From left to right in Figure~\ref{fig:cl0024},
  these galaxies are labeled \#362, \#374, and \#380 in the
  \cite{CZ01.1} catalog.

  As lenses, these three central galaxies can accurately reproduce the
  five observed arcs in Cl~0024$+$17, as shown in
  Figure~\ref{fig:cl0024}. \cite{OT04.1} note that all five arcs are
  images of a single background galaxy, and our results are consistent
  with that observation.

\end{itemize}

\section{Conclusions}
\label{conclusions}

We have modeled 11 clusters hosting gravitational arcs and observed
with HST as systems of dark matter halo lenses defined by elliptical
NFW density profiles. Given its position on the HST image, each lens is
completely defined by its scale convergence, scale radius,
ellipticity, and position angle. We use a minimization routine to vary
these parameters for each lens until the reproduced images match the
observed arcs in the cluster, and each observationally confirmed
family of arcs belongs to a unique source. We also require that the
predicted sources are compact. With this routine, we find the best-fit
scale convergence, scale radius, ellipticity, and position angle for
each lens in each cluster.

Our main results can be summarized as follows:

\begin{enumerate}

\item Each cluster in our sample is successfully modeled as a system
  of mass components with asymmetric NFW density profiles. The model
  produces images that correspond to observed arcs in the cluster, and
  the system of arc sources suggested by observations is reproduced.

\item The best-fit parameters to each modeled cluster fall within the
  range of reasonable values set by simulations. Also, our estimates
  of the lens masses are reasonable for large galaxies.

\item The accuracy of our minimization technique for finding the
  best-fit parameters to a system of NFW ellipsoids is verified by our
  fit to a simulated cluster with arcs. The cluster was simulated as
  an NFW ellipsoid, and we successfully reproduced the simulated
  cluster's parameters to within $4\%$.

\end{enumerate}

Clusters containing arcs have been commonly and successfully modeled
using mass components defined with spherical or ellipsoidal isothermal
density profiles. We show, however, that galaxy clusters can also be
convincingly modeled with NFW ellipsoids. The NFW profile is predicted
for dark matter halos in numerical simulations of cosmological
structure formation, and our result lends more credibility to NFW
profiles as models of actual observed galaxy clusters.

\acknowledgements J.M.C. acknowledges support by a National Science
Foundation Graduate Research Fellowship. J.M.C. thanks the Max Planck
Institute for Astrophysics and the Institute for Theoretical
Astrophysics at the University of Heidelberg for their warm
hospitality during her visits. This work was supported in part by the
\emph{Sonderforschungsbereich} 439 of the \emph{Deutsche
Forschungsgemeinschaft}.

\bibliographystyle{apj}
\bibliography{HST}

\end{document}